\documentclass[twocolumn,amsmath,amssymb,aip,pf]{revtex4}

\usepackage{graphicx}
\usepackage{bm}

\usepackage{amsfonts}
\usepackage{amsmath,bm}

\begin{document}

\title{Fourth-order modon in a rotating self-gravitating fluid}

\author{Volodymyr M. Lashkin}
\email{vlashkin62@gmail.com} \affiliation{Institute for Nuclear
Research, Pr. Nauki 47, Kyiv 03028, Ukraine} \affiliation{Space
Research Institute, Pr. Glushkova 40 k.4/1, Kyiv 03187, Ukraine}
\author{Oleg K. Cheremnykh}
\affiliation{Space Research Institute, Pr. Glushkova 40 k.4/1,
Kyiv 03187, Ukraine}

\begin{abstract}
We present a two-dimensional nonlinear equation to govern the
dynamics of disturbances in a rotating self-gravitating fluid. The
nonlinear term of the equation has the form of a Poisson bracket
(Jacobian), and the linear part contains, along with the
Laplacian, a biharmonic operator. A solution was found in the form
of a dipole vortex (modon). The solution and all its derivatives
up to the fourth order are continuous on the separatrix.
\end{abstract}

\maketitle

\section{Introduction}

The dynamics of self-gravitating systems (within the framework of
Newtonian theory) was first studied by Jeans \cite{Jeans1929},
where the instability of disturbances with wavelengths greater
than the so-called Jeans wavelength $\lambda_{J}$ was predicted.
It is believed that the Jeans instability is the source of the
emergence of large-scale structures in the Universe, and this
problem remains one of the most challenging in astrophysics. The
linear theory is valid, however, only for sufficiently small
(strictly speaking, infinitesimal) amplitudes of disturbances,
when nonlinear effects can be neglected. Nonlinear coherent
structures, such as solitons and vortices, in self-gravitating
fluids have been considered in a number of works. One-dimensional
solitons in such systems were studied in
\cite{Mikhailovskii1977,Yueh1981,Ono1994,Zhang1995,Zhang1998,
Gotz1988,Adams1994,Semelin2001,Verheest1997,Masood2010,Verheest2005},
and multidimensional structures were discussed in
\cite{Fridman1991,Jovanovich1990,Shukla1993,Zinzadze2000,
Pokhotelov1998,Shukla1995,Abrahamyan2020,Lashkin2023}. In a broad
sense, a soliton is a localized structure (not necessarily
one-dimensional) resulting from the balance of dispersion and
nonlinearity effects. Multidimensional solitons often turn out to
be unstable, and the most well-known phenomena in this case are
wave collapse and wave breaking
\cite{Kuznetsov1986,KuznetsovZakharov2000,Zakharov_UFN2012}. The
two-dimensional (2D) and three-dimensional (3D) structures in
\cite{Fridman1991,Jovanovich1990,Shukla1993,Zinzadze2000,
Pokhotelov1998,Shukla1995,Abrahamyan2020,Lashkin2023} were
considered under the assumption that the characteristic
frequencies are much lower than the rotation frequency of the
self-gravitating fluid. In geophysics, this corresponds to the
quasi-geostrophic approximation \cite{Pedlosky1987}. In
particular, in \cite{Fridman1991} a nonlinear equation governing
the dynamics of finite amplitude disturbances in a rotating
self-gravitating gas was obtained, and this equation is completely
similar to the well-known Charney equation in geophysics
\cite{Charney1948} for nonlinear Rossby waves in atmospheres of
rotating planets and in oceans (in plasma physics, this equation
is known as the Hasegawa-Mima equation \cite{Hasegawa1978}, and
the rotation frequency is replaced by the ion gyrofrequency in an
external magnetic field). The corresponding solution is the 2D
Larichev-Reznik dipole vortex in the form of a cyclone-anticyclone
pair (also called the Larichev-Reznik modon or Larichev-Reznik
soliton) \cite{Larichev1976a,Larichev1976b}. The term modon was
introduced by Stern \cite{Stern1975} to designate a steadily
moving localized vortex structure, and is widely used for
particular solutions of some type of equations with nonlinearity
in the form of Poisson brackets (mainly in geophysics and plasma
physics)
\cite{Flierl1987,Petviashvili_book1992,Berestov1979,Horton1983,Swaters1989,Ghil2002,Kizner2008,Zeitlin2022}.
A common property of these solutions is the presence of a circular
(in the 2D case) or spherical (3D case) cut (the separatrix) with
subsequent matching of the solution in the internal and external
regions. A remarkable property of the Larichev-Reznik modon is its
stability under head-on and overtaking collisions with zero-impact
parameter between the modons
\cite{Flierl1980,Makino1981,Williams1982}. Recent work
\cite{Lashkin2017} demonstrated elastic collisions even between
the 3D modons for the 3D version of the Hasegawa-Mima equation.

In most works on modons and their generalizations, the linear part
of the equations contains derivatives of no higher than second
order and, as a consequence, continuity of both the solution
itself and its derivatives no higher than the second order is
required. A few exceptions are, for example, works
\cite{Horton1986,Lakhin1987,Aburdjania1988PhysScr}, where
solutions in the form of dipole vortices were considered within
the framework of a system of two nonlinear equations for two
independent scalar functions, which is equivalent to an equation
with fourth-order derivatives. In the present paper, for the case
of a self-gravitating fluid, we present a nonlinear equation
including fourth-order derivatives in the linear part and a
nonlinear term in the form of Poisson brackets (Jacobian). We find
an analytical solution to this equation in the form of a dipole
vortex (modon). The solution and all its derivatives up to the
fourth order are continuous on the seperatrix.

The paper is organized as follows. In Sec. II, we present the
model equation for the rotating self-gravitating fluid. In Sec.
III, an analytical modon solution is found. Section IV concludes
the paper.

\section{ Model Equations}

We consider a self-gravitating fluid rotating with constant
angular velocity $\mathbf{\Omega}_{0}=\Omega_{0}\mathbf{\hat{z}}$
and with an equilibrium density $\rho_{0}$ in the plane
perpendicular to the $\mathbf{\hat{z}}$-axis. It is also assumed
that the characteristic frequencies of disturbances are small
compared to the rotation frequency. A weak inhomogeneity of the
equilibrium density in the radial direction with a characteristic
inhomogeneity length $L$ (all characteristic scales of
perturbations are much larger than $L$) is assumed, and a local
Cartesian coordinate system is used ($x$ corresponds to the radial
coordinate $r$ and $y$ corresponds to the polar angle $\varphi$).
Based on the momentum and continuity fluid equations for the
self-gravitating rotating isothermal gas and the Poisson equation
for the gravitational potential, the following system of coupled
3D nonlinear equations for the perturbed gravitational potential
$\psi$ and the $z$-component of the fluid velocity $v_{z}$ was
obtained in a recent paper \cite{Lashkin2023},
\begin{equation}
\label{main2} \frac{\partial \Phi}{\partial
t}-\frac{\omega_{0}^{2}}{2\Omega_{0}L} \frac{\partial
\Pi}{\partial y}
+\frac{1}{2\Omega_{0}}\{\Pi,\Phi\}+\omega_{0}^{2}\frac{\partial
v_{z}}{\partial z}=0,
\end{equation}
\begin{equation}
\label{main_z} \frac{\partial v_{z}}{\partial
t}+\frac{1}{2\Omega_{0}}\left\{\Pi,v_{z}\right\}+\frac{\partial
\Pi}{\partial z}=0,
\end{equation}
where
\begin{equation}
\label{Phi}
\Phi=\Delta\psi-\frac{\omega_{0}^{2}}{4\Omega_{0}^{2}}\Delta_{\perp}\Pi
, \quad \Pi=\psi+\frac{c_{s}^{2}} {\omega_{0}^{2}}\Delta\psi,
\end{equation}
being $\Delta=\partial^{2}/\partial x^{2}+\partial^{2}/\partial
y^{2}+\partial^{2}/\partial z^{2}$ and
$\Delta_{\perp}=\partial^{2}/\partial x^{2}+\partial^{2}/\partial
y^{2}$ are the 3D and 2D (transverse) Laplacian respectively. Here
$\omega_{0}$ is the Jeans frequency defined by
$\omega_{0}=2\sqrt{\pi G\rho_{0}}$, where $G$ is the gravitational
constant, and $c_{s}$ is the speed of sound in an adiabatic medium
of uniform density $\rho_{0}$. The nonlinear terms in Eqs.
(\ref{main2}) and (\ref{main_z}) have been written in the form of
the Poisson bracket (Jacobian) defined as
\begin{equation}
\{f,g\}=\frac{\partial f}{\partial x}\frac{\partial g}{\partial
y}-\frac{\partial f}{\partial y}\frac{\partial g}{\partial x}.
\end{equation}
We consider the 2D case and neglect the motions along the
$z$-axis, which is valid provided
\begin{equation}
\frac{\partial/\partial t}{\Omega_{0}}\gg
\frac{l_{y}}{l_{z}^{2}}L,
\end{equation}
where $l_{y}$ and $l_{z}$ are characteristic lengths of
disturbances along the $y$-axis and $z$-axis, respectively. One
can verify that this is equivalent to neglecting the interaction
with the acoustic wave in \cite{Lashkin2023}. From Eqs.
(\ref{main2}) and (\ref{main_z}) we then obtain one equation for
the potential $\psi$,
\begin{equation}
\label{main4} \frac{\partial\Phi}{\partial
t}-\frac{\omega_{0}^{2}}{2\Omega_{0}L}\frac{\partial\Pi}{\partial
y}+\frac{1}{2\Omega_{0}}\{\Pi,\Phi\}=0,
\end{equation}
where in Eq. (\ref{Phi}) the 3D Laplacian is replaced by a
two-dimensional one,  $\Delta\rightarrow\Delta_{\perp}$. In the
linear approximation, taking $\psi (\mathbf{r}_{\perp},t)\sim \exp
(i\mathbf{k}_{\perp}\cdot\mathbf{r}_{\perp}-i\omega t)$, where
$\omega$ and $\mathbf{k}_{\perp}=(k_{x},k_{y})$ are the frequency
and transverse wave vector respectively, Eq. (\ref{main4}) yields
the dispersion relation
\begin{equation}
\label{branch1}
\omega=\frac{k_{y}c_{s}^{2}}{2\Omega_{0}L[k^{2}_{\perp}/(k^{2}_{\perp}-k^{2}_{J})
+k^{2}_{\perp}c_{s}^{2}/(4\Omega_{0}^{2})]},
\end{equation}
where $k_{\perp}^{2}=k_{x}^{2}+k_{y}^{2}$, $k_{J}=1/\lambda_{J}$,
and $\lambda_{J}=c_{s}/\omega_{0}$ is the Jeans length. Note that
the solution in the form of a monochromatic plane wave is also a
solution to Eq. (\ref{main4}) since in this case the nonlinearity
in the form of a Poisson bracket vanishes identically. Introducing
the variables $t^{'}$, $\mathbf{r}_{\perp}^{'}$, and $\psi^{'}$ by
\begin{equation}
t=\frac{t^{\prime}}{2\Omega_{0}},\quad
\mathbf{r}_{\perp}=\lambda_{J}\mathbf{r}_{\perp}^{\prime} ,\quad
\psi=\frac{4\Omega_{0}^{2}c_{s}^{2}}{\omega_{0}^{2}}\psi^{\prime},
\end{equation}
we rewrite Eq. (\ref{main4}) in the dimensionless form (primes
have been omitted),
\begin{gather}
\frac{\partial}{\partial
t}(\alpha\Delta_{\perp}\psi-\Delta_{\perp}^{2}\psi)-v_{\ast}\frac{\partial}{\partial
y}(\psi+\Delta_{\perp}\psi)
 \nonumber \\
+\{\psi+\Delta_{\perp}\psi,\alpha\Delta_{\perp}\psi-\Delta_{\perp}^{2}\psi\}=0,
\label{main5}
\end{gather}
where $\alpha=4\Omega_{0}^{2}/\omega_{0}^{2}-1$, and
$v_{\ast}=\lambda_{J}/L$. Note that our basic equation
(\ref{main5}) contains a biharmonic operator, that is, in
particular, fourth-order spatial derivatives. Equation
(\ref{main5}) can be written as
\begin{equation}
\label{main6} \frac{\partial \Gamma}{\partial
t}+\{\psi+\Delta_{\perp}\psi,\Gamma\}=0,
\end{equation}
where
$\Gamma=\Delta_{\perp}^{2}\psi-\alpha\Delta_{\perp}\psi-v_{\ast}x$
can be treated as a generalized vorticity, and then Eq.
(\ref{main6}) describes the generalized vorticity convection in an
incompressible velocity field
$\mathbf{v}_{D}=[\hat{\mathbf{z}}\times\nabla_{\perp}
(\psi+\Delta_{\perp}\psi)]$ with $d\Gamma/dt=0$, where
$d/dt=\partial/\partial t+\mathbf{v}_{D}\cdot\nabla_{\perp}$ is
the convective derivative. Conservation of the generalized
vorticity $\Gamma$ along the streamlines implies that Eq.
(\ref{main5}) has an infinite set of integrals of motion (the so
called Casimir invariants),
\begin{equation}
\label{integral1} \int f(\Gamma)\,d^{2}\mathbf{r},
\end{equation}
where $f$ is an arbitrary function. Other integrals of motion are
\begin{equation}
\label{integral2} \int
(\psi+\Delta_{\perp}\psi)\Gamma\,d^{2}\mathbf{r}, \quad \int
x\Gamma\,d^{2}\mathbf{r}, \quad  \int
(y+v_{\ast}t)\Gamma\,d^{2}\mathbf{r}.
\end{equation}

\section{Modon solution}

We look for stationary traveling wave solutions of Eq.
(\ref{main6}) of the form
\begin{equation}
\psi (x,y,t)=\psi (x,y^{\prime},t), \quad y^{\prime}=y-ut,
\end{equation}
where $u$ is the velocity of propagation in the $y$ direction and
in the following we omit the prime. Then Eq. (\ref{main6}) becomes
\begin{equation}
-u\frac{\partial\Gamma}{\partial
y}+\{\psi+\Delta_{\perp}\psi,\Gamma\}=0,
\end{equation}
and can be written as a single Poisson bracket relation
\begin{equation}
\{\psi+\Delta_{\perp}\psi-ux,\Gamma\}=0.
\end{equation}
It follows that the functions in the bracket are dependent, and
then
\begin{equation}
\label{F} \Gamma=F(\psi+\Delta_{\perp}\psi-ux),
\end{equation}
that is
\begin{equation}
\label{F1}
\Delta_{\perp}^{2}\psi-\alpha\Delta_{\perp}\psi-v_{\ast}x=F(\psi+\Delta_{\perp}\psi-ux),
\end{equation}
where $F$ is an arbitrary function. We are looking localized
solutions, that is $\psi\rightarrow 0$ as $x,y\rightarrow\infty$.
In addition to localization at infinity, the solution must also be
finite at zero. Following the well-known procedure for finding the
modon-type solutions \cite{Larichev1976a}, we solve Eq. (\ref{F1})
in polar coordinates ($r,\theta$) in two regions (inner and
external) separated by a circle of radius $a$ (the separatrix).
The separatrix separates trapped and untrapped fluid. In the
external region $r>a$, taking the $y\rightarrow\infty$ limit of
Eq. (\ref{F1}), one can see that the function $F$ must be linear,
$F(\xi)=c_{1}\xi$, where $c_{1}$ is a constant, and Eq. (\ref{F1})
becomes
\begin{equation}
\label{linear-ex}
\Delta_{\perp}^{2}\psi-\alpha\Delta_{\perp}\psi-v_{\ast}x=c_{1}(\psi+\Delta_{\perp}\psi-ux).
\end{equation}
The localization condition with respect to $x$ uniquely determines
the constant $c_{1}=v_{\ast}/u$. Then from Eq. (\ref{linear-ex})
we have
\begin{equation}
\label{linear-ex1}
\Delta_{\perp}^{2}\psi-\left(\alpha+\frac{v_{\ast}}{u}\right)\Delta_{\perp}\psi-\frac{v_{\ast}}{u}\psi=0.
\end{equation}
In the inner region $r<a$ we choose the function $F$ in Eq.
(\ref{F1}) to also be linear, $F(\xi)=c_{2}\xi$, where $c_{2}$ is
a constant, and Eq. (\ref{F1}) becomes
\begin{equation}
\label{linear-in}
\Delta_{\perp}^{2}\psi-\alpha\Delta_{\perp}\psi-v_{\ast}x=c_{2}(\psi+\Delta_{\perp}\psi-ux),
\end{equation}
which we rewrite as
\begin{equation}
\label{linear-in1}
\Delta_{\perp}^{2}\psi-\left(\alpha+c_{2}\right)\Delta_{\perp}\psi-c_{2}\psi-(v_{\ast}-u)x=0.
\end{equation}
We are looking for solutions to Eqs. (\ref{linear-ex1}) and
(\ref{linear-in1}) of the form
\begin{equation}
\label{radial-form} \psi(r,\theta)=\Phi (r) \cos \theta ,
\end{equation}
with the boundary conditions $\Phi (r)\rightarrow 0$ as
$r\rightarrow\infty$, and $\Phi (r)$ is regular at $r=0$.
Substituting Eq. (\ref{radial-form}) into Eqs. (\ref{linear-ex1})
and (\ref{linear-in1}) gives
\begin{equation}
\label{rad-ex1}
\Delta_{(r)}^{2}\Phi-\left(\alpha+\frac{v_{\ast}}{u}\right)\Delta_{(r)}\Phi-\frac{v_{\ast}}{u}\Phi=0,
\,\, (r > a),
\end{equation}
and
\begin{equation}
\label{rad-in1}
\Delta_{(r)}^{2}\Phi-\left(\alpha+c_{2}\right)\Delta_{(r)}\Phi-c_{2}\Phi-(v_{\ast}-u)r=0,\,\,
(r < a),
\end{equation}
where
\begin{equation}
\Delta_{(r)}=\frac{d^{2}}{dr^{2}}+\frac{1}{r}\frac{d}{dr}-\frac{1}{r^{2}}.
\end{equation}
In the external region $r>a$, the general solution of Eq.
(\ref{rad-ex1}) localized at infinity has the form
\begin{equation}
\label{psi-ex}
\Phi_{\mathrm{ex}}=AK_{1}(\lambda_{1}r)+BK_{1}(\lambda_{2}r),
\end{equation}
where $K_{n}$ is the $n$-order McDonald function, $A$ and $B$ are
the integration constants, and
\begin{equation}
\lambda_{1,2}^{2}=\frac{1}{2}\left[\alpha+\frac{v_{\ast}}{u}\pm
\sqrt{\left(\alpha+\frac{v_{\ast}}{u}\right)^{2}+\frac{4v_{\ast}}{u}}\right].
\end{equation}
In order for $\lambda_{1,2}$ to be real, the restrictions
\begin{equation}
\label{cond1} u<0, \quad \alpha+\frac{v_{\ast}}{u}>0, \quad
\left(\alpha+\frac{v_{\ast}}{u}\right)^{2}+\frac{4v_{\ast}}{u}>0
\end{equation}
must be met. The first of the conditions (\ref{cond1}) means that
the solution travels only in the negative direction of the $y$
axis. From Eq. (\ref{cond1}), it also follows the condition for
the ratio of the angular velocity of the system $\Omega_{0}$ and
the Jeans frequency $\omega_{0}$,
\begin{equation}
\frac{4\Omega_{0}^{2}}{\omega_{0}^{2}}>1+\frac{v_{\ast}}{|u|}+2\sqrt{\frac{v_{\ast}}{|u|}}.
\end{equation}
Let us now consider Eq. (\ref{rad-in1}) for the inner region
$r<a$. A general solution of Eq. (\ref{rad-in1}) is the sum of the
general solution of the corresponding homogeneous equation and the
particular solution $-(v_{\ast}-u)r/c_{2}$. Depending on the signs
and relationships between the coefficients of Eq. (\ref{rad-in1}),
there are three different general solutions of this equation
finite at zero,
\begin{gather}
\Phi_{\mathrm{in}}=CJ_{1}(\mu_{1}r)+DI_{1}(\mu_{2}r)-\frac{v_{\ast}+|u|}{c_{2}}r,
\label{variant1}
\\
\Phi_{\mathrm{in}}=CJ_{1}(\mu_{3}r)+DJ_{1}(\mu_{4}r)-\frac{v_{\ast}+|u|}{c_{2}}r,
\label{variant2}
\\
\Phi_{\mathrm{in}}=CI_{1}(\mu_{5}r)+DI_{1}(\mu_{6}r)-\frac{v_{\ast}+|u|}{c_{2}}r,
\label{variant3}
\end{gather}
where
\begin{gather}
\mu_{1,2}^{2}=\frac{1}{2}\left[
\sqrt{\left(\alpha+c_{2}\right)^{2}+4c_{2}}\pm
\left(\alpha+c_{2}\right)\right], \label{roots1}
\\
\mu_{3,4}^{2}=\frac{1}{2}\left[-\left(\alpha+c_{2}\right)\pm
\sqrt{\left(\alpha+c_{2}\right)^{2}+4c_{2}}\right], \label{roots2}
\\
\mu_{5,6}^{2}=\frac{1}{2}\left[\left(\alpha+c_{2}\right)\pm
\sqrt{\left(\alpha+c_{2}\right)^{2}+4c_{2}}\right]. \label{roots3}
\end{gather}
Here $J_{n}$ and $I_{n}$ are the $n$-order Bessel and modified
Bessel functions of the first kind, respectively, $C$ and $D$ are
the arbitrary integration constants. The reality of $\mu_{1,2}$,
$\mu_{3,4}$ and $\mu_{5,6}$ imposes the conditions,
\begin{gather}
c_{2}>0, \quad \mbox{for solution } (\ref{variant1}), \label{con1}
\\
(\alpha+c_{2})<0 \quad \mbox{and} \quad
(\alpha+c_{2})^{2}>-4c_{2}>0, \nonumber
\\ \quad \mbox{for solution } (\ref{variant2}), \label{con2}
\\
(\alpha+c_{2})>0 \quad \mbox{and} \quad
(\alpha+c_{2})^{2}>-4c_{2}>0, \nonumber
\\ \quad \mbox{for solution } (\ref{variant3}). \label{con3}
\end{gather}
Conditions (\ref{con2}) and (\ref{con3}) imply, in addition to Eq.
(\ref{cond1}), additional restrictions on $\alpha$. In the present
paper we restrict ourselves to the case of solution
(\ref{variant1}).

Just as for the Larichev-Reznik modon and its generalizations, the
exterior and interior solutions (\ref{psi-ex}) and
(\ref{variant1}) are matched at the separatrix $r=a$ so as to
guarantee the continuity of physically meaningful variable $\psi$
and its first derivatives. We require that all derivatives of the
function $\psi$ with respect to the radial coordinate $r$ be
continuous up to the fourth order. Since the angular dependence in
Eqs. (\ref{psi-ex}) and (\ref{variant1}) is contained only as a
multiplier, the matching conditions are written as
\begin{gather}
\psi_{\mathrm{ex}}\mid_{r=a}=\psi_{\mathrm{in}}\mid_{r=a},
\label{continuity1} \\
\nabla_{\perp}\psi_{\mathrm{ex}}\mid_{r=a}=\nabla_{\perp}\psi_{\mathrm{in}}\mid_{r=a},
\label{continuity2}
\\
\label{continuity3}
\Delta_{\perp}\psi_{\mathrm{ex}}\mid_{r=a}=\Delta_{\perp}\psi_{\mathrm{in}}\mid_{r=a}
\\
\nabla_{\perp}\Delta_{\perp}\psi_{\mathrm{ex}}\mid_{r=a}=\nabla_{\perp}\Delta_{\perp}\psi_{\mathrm{in}}\mid_{r=a},
\label{continuity4}
\\
\Delta^{2}_{\perp}\psi_{\mathrm{ex}}\mid_{r=a}=\Delta^{2}_{\perp}\psi_{\mathrm{in}}\mid_{r=a}.
\label{continuity5}
\end{gather}
Substituting Eqs. (\ref{psi-ex}) and (\ref{variant1}) into Eqs.
(\ref{continuity1})-(\ref{continuity5}) we obtain a system of
equations that determine the unknown coefficients $A$, $B$, $C$,
$D$, and constant $c_{2}$,
\begin{gather}
AK_{11}+BK_{12}-CJ_{11}-DI_{12} =f,
\label{e1} \\
A\lambda_{1}K_{21}+B\lambda_{2}K_{22}-C\mu_{1}J_{21}
+D\mu_{2}I_{22} =0, \label{e2}
\\
A\lambda_{1}^{2}K_{11}+B\lambda_{2}^{2}K_{12} +C\mu_{1}^{2}J_{11}
-D\mu_{2}^{2}I_{12}=0 , \label{e3}
\\
A\lambda_{1}^{3}K_{21}+B\lambda_{2}^{3}K_{22} +C\mu_{1}^{3}J_{21}
+D\mu_{2}^{3}I_{22}=0 \label{e4}
\\
A\lambda_{1}^{4}K_{11}+B\lambda_{2}^{4}K_{12} -C\mu_{1}^{4}J_{11}
-D\mu_{2}^{4}I_{12}=0, \label{e5}
\end{gather}
where, for brevity, we have introduced the notations
$K_{11}=K_{1}(\lambda_{1}a)$,$K_{12}=K_{1}(\lambda_{2}a)$,$K_{21}=K_{2}(\lambda_{1}a)$,
$K_{22}=K_{2}(\lambda_{2}a)$,$J_{11}=J_{1}(\mu_{1}a)$,$J_{21}=J_{2}(\mu_{1}a)$,
$I_{12}=I_{1}(\mu_{2}a)$,$I_{22}=I_{2}(\mu_{2}a)$, and
$f=-(v_{\ast}+|u|)a/c_{2}$. Equation (\ref{e2}) is obtained as a
result of a combination of Eqs. (\ref{continuity1}) and
(\ref{continuity2}), similarly Eq. (\ref{e4}) is a consequence of
Eqs. (\ref{continuity3}) and (\ref{continuity4}). From the system
(\ref{e1})-(\ref{e4}) we find
\begin{equation}
\label{ABCD} A=\frac{f\mathcal{A}}{G}, \quad
B=\frac{f\mathcal{B}}{G}, \quad C=\frac{f\mathcal{C}}{G}, \quad
D=\frac{f\mathcal{D}}{G},
\end{equation}
where $\mathcal{A}$, $\mathcal{B}$, $\mathcal{C}$, $\mathcal{D}$,
and $G$ are given in the Appendix. The unknown coefficient $c_{2}$
could be determined from the remaining equation (\ref{e5}). This
transcendental equation, however, is extremely cumbersome (recall
that $c_{2}$ is contained in $f$, $\mu_{1}$ and $\mu_{2}$).
Instead, we obtain a simpler (albeit also cumbersome)
transcendental equation for $c_{2}$ by setting the determinant of
the homogeneous system (\ref{e2})-(\ref{e5}) to zero. The
resulting equation is
\begin{gather}
(\lambda_{2}^{2}-\lambda_{1}^{2})(\mu_{1}^{2}+\mu_{2}^{2})\left[\lambda_{1}\lambda_{2}
K_{11}K_{12}J_{21}I_{22} \right.
\nonumber \\
\left. +\mu_{1}\mu_{2}
K_{21}K_{22}J_{11}I_{12}\right]+(\lambda_{1}^{2}+\mu_{1}^{2})(\lambda_{2}^{2}-\mu_{2}^{2})
\nonumber \\
\times\left[\lambda_{1}\mu_{1}K_{11}K_{22}J_{11}I_{22}+\lambda_{2}\mu_{2}K_{12}K_{21}J_{21}I_{12}\right]
\nonumber \\
-(\lambda_{1}^{2}-\mu_{2}^{2})(\lambda_{2}^{2}+\mu_{1}^{2})\left[\lambda_{1}\mu_{2}K_{11}K_{22}J_{21}I_{12}
\right.
\nonumber \\
\left. +\lambda_{2}\mu_{1}K_{12}K_{21}J_{11}I_{22}\right]=0 .
 \label{transcend}
\end{gather}
Thus, the solution to Eq. (\ref{main5}) in the form of a dipole
vortex soliton (modon) is
\begin{equation}
\label{psi-full} \psi (r,\theta)=\cos\theta \!
\begin{cases}
\displaystyle CJ_{1}(\mu_{1}r)\!+\!DI_{1}(\mu_{2}r)-\frac{v_{\ast}\!+\!|u|}{c_{2}}r,& \! r \leqslant a, \\
\displaystyle AK_{1}(\lambda_{1}r)\!+\!BK_{1}(\lambda_{2}r),& \! r
\geqslant a.
\end{cases}
\end{equation}
\begin{figure}
\includegraphics[width=3.2in]{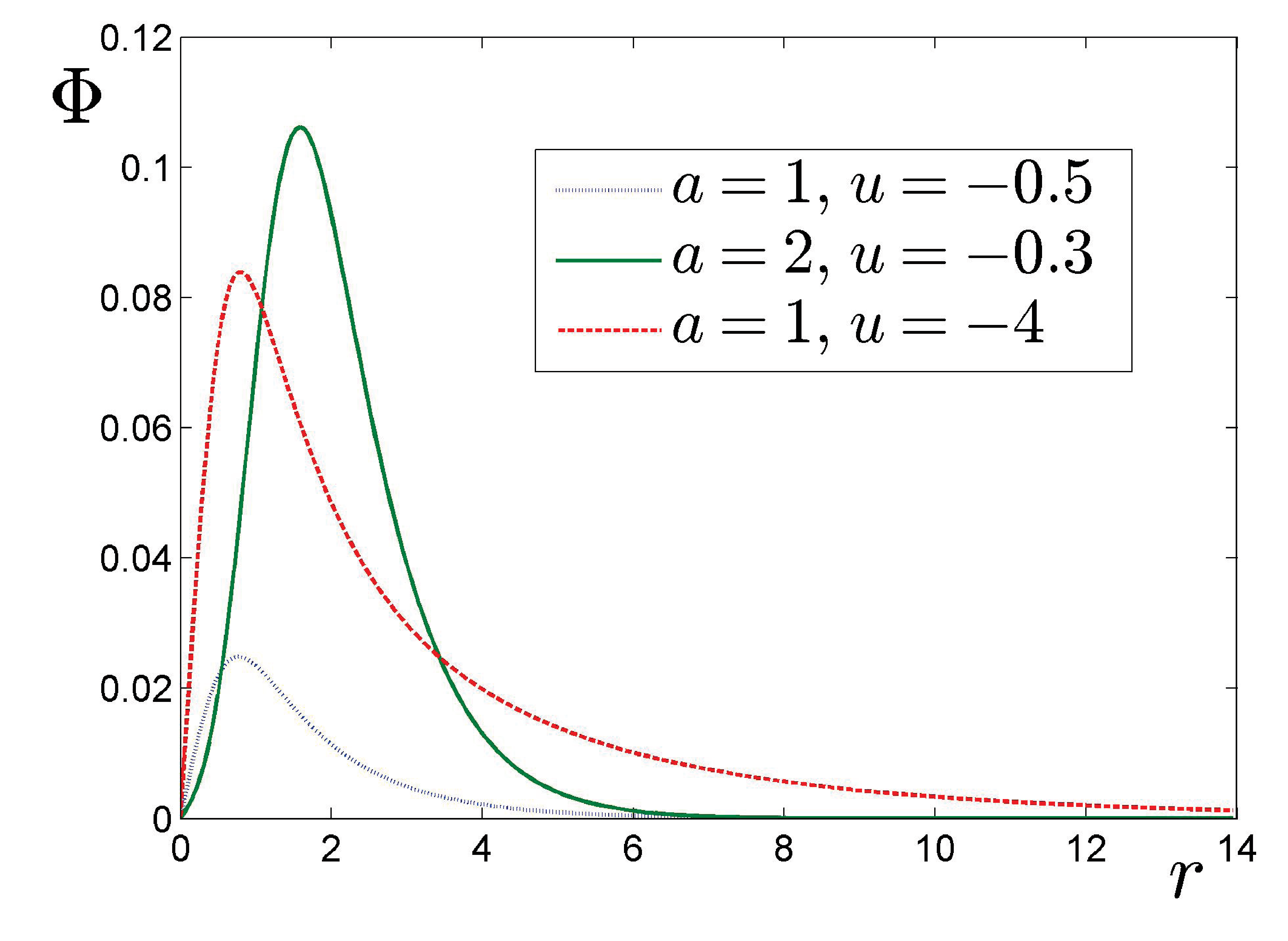}
\caption{\label{fig1} Radial profiles of the modon solution
$\Phi(r)$ for different $a$ (cut radius) and $u$ (modon velocity)
with $v_{\ast}=1$ and $\alpha=7$. }
\end{figure}

\begin{figure}
\includegraphics[width=3.2in]{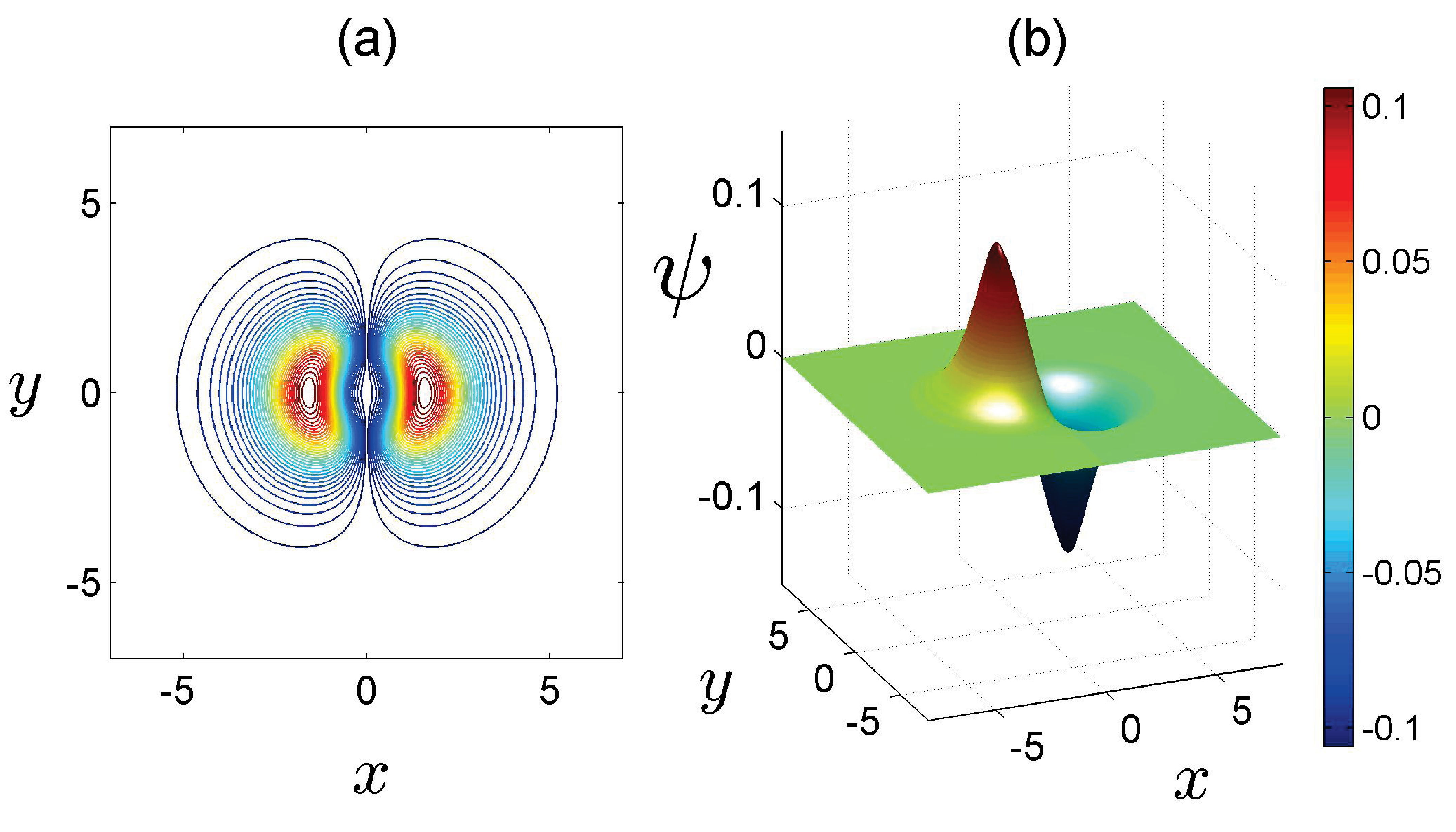}
\caption{\label{fig2} The modon (\ref{psi-full}) with the
parameters $a=2$ (cut radius), $u=-0.3$ (velocity) for
$v_{\ast}=1$ and $\alpha=7$. (a) Streamlines $|\psi|$ in the
$(x,y)$ plane; (b) modon solution $\psi$. }
\end{figure}

\begin{figure}
\includegraphics[width=3.2in]{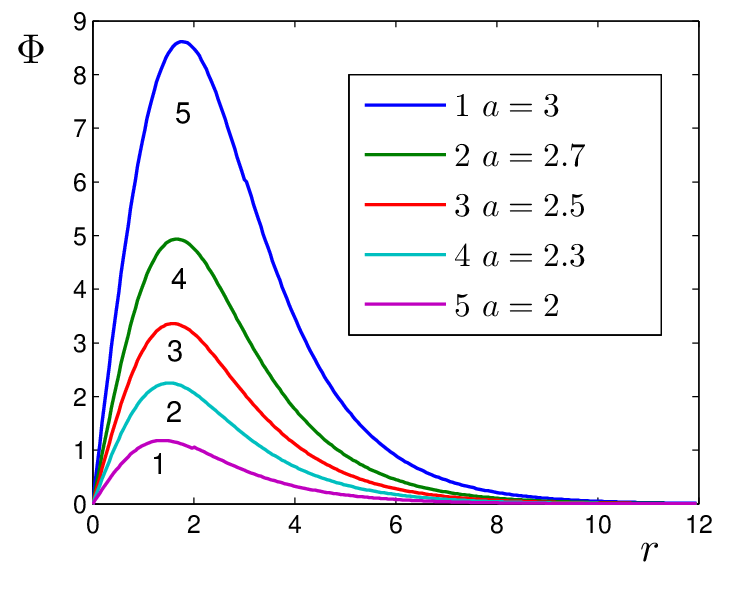}
\caption{\label{fig3} Radial profiles of the modon solution
$\Phi(r)$ for different $a$ (cut radius) at the fixed modon
velocity $u=-2$ with $v_{\ast}=1$ and $\alpha=2$. }
\end{figure}

\begin{figure}
\includegraphics[width=3.2in]{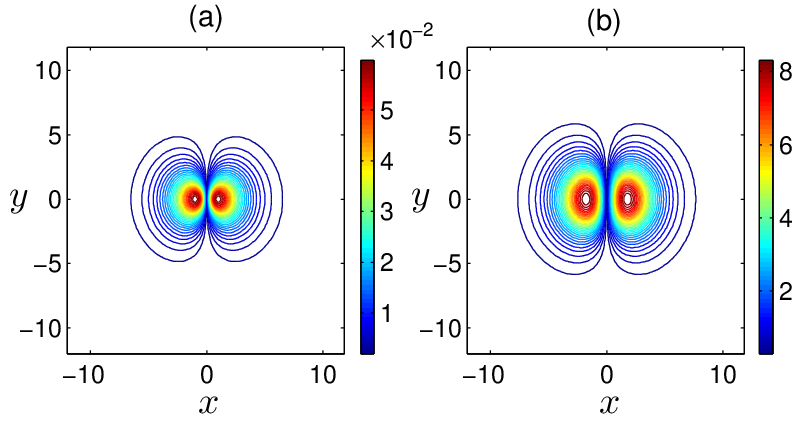}
\caption{\label{fig4} Streamlines $|\psi|$ in the $(x,y)$ plane at
the fixed modon velocity $u=-2$ and different $a$ with
$v_{\ast}=1$ and $\alpha=2$. a) $a=1$; b) $a=3$.}
\end{figure}

\begin{figure}
\includegraphics[width=3.2in]{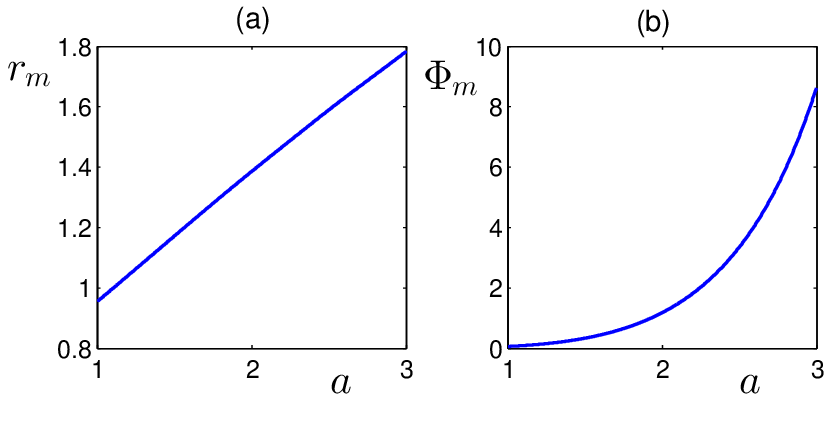}
\caption{\label{fig5} (a) Dependence of the radial position of the
maximum of the radial part (amplitude) of the modon $r_{m}$ on the
separatrix radius $a$; (b) dependence of the amplitude of the
radial part of the modon $\Phi_{m}$ on the separatrix radius $a$;
the parameters: $u=-2$, $v_{\ast}=1$ and $\alpha=2$.}
\end{figure}
The coefficients $A$, $B$, $C$, and $D$ in Eqs. (\ref{psi-ex}) and
(\ref{variant1}) are given by Eq. (\ref{ABCD}), and the constant
$c_{2}$ is determined from Eq. (\ref{transcend}). Note that the
transcendental equation (\ref{transcend}) has an infinite set of
roots $c_{2,n}$, $n=1,2 \dots$ for each fixed pair of values of
the free parameters $u$ (velocity) and $a$ (separatrix radius).
The solution with $n = 1$ (the ground-state modon) has no radial
nodes. Further, we consider only the ground-state modon with
$c_{2,1}\equiv c_{2}$. Equation (\ref{transcend}) for $c_{2}$ can
be readily solved, and then the coefficients $A$, $B$, $C$, and
$D$ are uniquely determined, and with them the exact solution
(\ref{psi-full}) is found that is continuous on the separatrix
$r=a$ up to the fourth derivative. Radial profiles of the modon
solution $\Phi (r)$ for different $a$ (cut radius) and $u$ (modon
velocity) are plotted in Fig.~1. In all these cases shown, we fix
the parameters $v_{\ast}=1$ and $\alpha=7$. Note that the point
$r_{m}$ corresponding to the maximum of the function $\Phi (r)$ is
always located to the left of the cut radius point, $r_{m}<a$
($\Phi (r)$ has an extremum only in the internal region).
Streamlines $|\psi|$ in the $(x,y)$ plane and the modon solution
(\ref{psi-full}) are shown in Fig.~2. We also plotted in Fig.~3
the radial parts of the modon solution for several values of the
separatrix radius $a$ at a fixed modon velocity $u=-2$
($v_{\ast}=1$ and $\alpha=2$), and streamlines $|\psi|$ in the
$(x,y)$ plane for two different $a$ (Fig.~4). The maximum
amplitude of the modon rapidly decreases with decreasing radius of
the separatrix $a$, for example, for $a=3$ the maximum amplitude
is $\Phi_{m}\approx 8.653$, whereas already for $a=1$ we have
$\Phi_{m}\approx 0.062$. The dependence of the position $r_{m}$ of
the maximum amplitude (that is, the actual size of the modon) on
the radius of the separtrix turns out to be almost linear. The
dependences of $r_{m}$ and  maximum amplitude of the modon
$\Phi_{m}$ on the radius of the separatrix $a$ at a fixed $u=-2$
are presented in Fig.~5.

In the limit $u\rightarrow 0$ with $a$ fixed from Eq.
(\ref{linear-ex}) one can see that the solution in the outer
region must be equal to zero. In Eqs. (\ref{e1})-(\ref{e5}) we
then have $A$=0 and $B=0$, and to determine $c_{2}$ we find
\begin{equation}
\label{trans2}  \mu_{1} J_{11}I_{22}=\mu_{2} J_{21}I_{12}.
\end{equation}
The coefficients $C$ and $D$ in this case have the form
\begin{gather}
C=\frac{v_{\ast} a\mu_{2}I_{22}}{c_{2}(\mu_{1}
J_{21}I_{12}+\mu_{2} J_{11}I_{22})}, \label{CC}
\\
D=\frac{v_{\ast} a\mu_{1}J_{21}}{c_{2}(\mu_{1}
J_{21}I_{12}+\mu_{2} J_{11}I_{22})}. \label{DD}
\end{gather}
The solution is
\begin{equation}
\label{psi-Stern} \psi (r,\theta)=
\begin{cases}
\displaystyle \left[CJ_{1}(\mu_{1}r)+DI_{1}(\mu_{2}r)-v_{\ast}r/c_{2}\right]\cos\theta ,&  r \leqslant a, \\
\displaystyle 0 ,& r \geqslant a .
\end{cases}
\end{equation}
Here the nonlinearity acts to completely screen the modon for
$r>a$, similar to the modon obtained by Stern \cite{Stern1975}.
Note that in this case the solution and the first and second
derivatives are continuous on the separatrix.

The problem of stability of the found modon is an open and
difficult question (as indeed for second-order modons). Here we
only note that despite the impressive demonstrations of fully
elastic collisions of Larichev-Reznik modons
\cite{Flierl1980,Makino1981,Williams1982} and also
three-dimensional modons in \cite{Lashkin2017}, the question of
stability with respect to arbitrary perturbations is still unclear
(see, e. g., \cite{Swaters2004} and references therein). In
particular, in \cite{Swaters2004} it was indicated that the
Larichev-Reznik modon moving in the negative $x$-direction (in our
notation this corresponds to the positive $y$-direction) always
undergoes the so-called tilt instability \cite{Nycander1992},
which manifests itself under an infinitesimal small perturbation
of the direction of modon propagation. Since in our case the found
fourth-order modon can only move in the negative $y$-direction,
then, at least if we make analogies with the Larichev-Reznik
modon, one can expect that the modon we found will not suffer from
the tilt instability.

\section{Conclusion}

We have obtained a two-dimensional nonlinear equation describing
the dynamics of disturbances in a rotating self-gravitating fluid
under the assumption that the rotation frequency is much higher
than the characteristic frequencies of disturbances. The resulting
equation contains, in particular, a biharmonic operator in both
the linear and nonlinear parts. The nonlinear term is a Poisson
bracket (Jacobian). We have obtained a solution in the form of a
solitary dipole vortex (modon) using a procedure similar to the
Larichev-Reznik method, that is, introducing a cut on a plane
(separatrix). The resulting modon solution and all its derivatives
up to the fourth order are continuous on the separatrix.

It should be noted that the well-known generalization of the
Larichev-Reznik solution \cite{Flierl1980} contains, in addition
to the dipole part (carrier), also a radially symmetric monopole
part (the so-called rider), the amplitude of which is a free
parameter and, generally speaking, can significantly exceed the
amplitude of the dipole part, although the monopole part does not
exist without the dipole one. In this case, the second derivative
(and vorticity) has a constant jump on the separatrix. The
magnitude of the jump determines the amplitude of the monopole
part. In the present paper, we have restricted ourselves to
solution (\ref{psi-full}) without monopole part, although
obtaining the corresponding solution with the monopole part can
apparently be carried out without difficulty. In this case,
however, the fourth derivative of the solution will have a
discontinuity.

The found fourth-order modon in the form of a cyclone-anticyclone
dipole pair in a self-gravitating fluid can apparently exist in
astrophysical objects. For example, the observation of the Mrk 266
galaxy with two nuclei, rotating in the opposite direction was
reported in \cite{Abrahamyan2020}.

A very interesting task is the numerical simulation of the
collision between the found fourth-order modons. We plan to
implement this in the future.

\section*{ AUTHOR DECLARATIONS}

\subsection*{Conflict of Interest}

The authors have no conflicts to disclose.

\subsection*{Author Contributions}

\textbf{V. M. Lashkin}: Conceptualization (equal), Methodology
(equal), Validation (equal), Formal analysis (equal),
Investigation (equal). \textbf{O. K. Cheremnykh}:
Conceptualization (equal), Methodology (equal), Validation
(equal), Formal analysis (equal), Investigation (equal).

\section*{ DATA AVAILABILITY}

No data are associated with this manuscript.

\section{Appendix}

This appendix provides the explicit expressions for $\mathcal{A}$,
$\mathcal{B}$, $\mathcal{C}$, $\mathcal{D}$, and $G$ in Eq.
(\ref{ABCD}),
\begin{gather}
\mathcal{A}=\lambda_{2}\mu_{1}\mu_{2}\left[\lambda_{2}(\mu_{1}^{2}
+\mu_{2}^{2})K_{12}J_{21}I_{22}
\right. \nonumber \\
\left. -
\mu_{1}(\lambda_{2}^{2}-\mu_{2}^{2})K_{22}J_{11}I_{22}
+\mu_{2}(\lambda_{2}^{2}+\mu_{1}^{2})K_{22}J_{21}I_{12}\right],
\label{A}
\\
\mathcal{B}=\lambda_{1}\mu_{1}\mu_{2}\left[\mu_{1}(\lambda_{1}^{2}-\mu_{2}^{2})K_{21}J_{11}I_{22}
\right. \nonumber \\
\left. -\lambda_{1}(\mu_{1}^{2}+\mu_{2}^{2})K_{11}J_{21}I_{22}
-\mu_{2}(\lambda_{1}^{2}+\mu_{1}^{2} )K_{21}J_{21}I_{12}\right],
\label{B}
\\
\mathcal{C}=\lambda_{1}\lambda_{2}\mu_{2}
\left[\lambda_{1}(\lambda_{2}^{2}-\mu_{2}^{2})K_{11}K_{22}I_{22}
\right. \nonumber \\
\left. -
\lambda_{2}(\lambda_{1}^{2}-\mu_{2}^{2})K_{12}K_{21}I_{22}-\mu_{2}
(\lambda_{1}^{2}-\lambda_{2}^{2})K_{21}K_{22}I_{12}\right],
\label{C}
\\
\mathcal{D}=\lambda_{1}\lambda_{2}\mu_{1}
\left[\lambda_{1}(\lambda_{2}^{2}+\mu_{1}^{2})K_{11}K_{22}J_{21}
\right. \nonumber \\
\left.
-\lambda_{2}(\lambda_{1}^{2}+\mu_{1}^{2})K_{12}K_{21}J_{21}-\mu_{1}
(\lambda_{1}^{2}-\lambda_{2}^{2})K_{21}K_{22}J_{11}\right],
\label{D}
\end{gather}
and
\begin{gather}
G=(\lambda_{1}^{2}-\lambda_{2}^{2})(\mu_{1}^{2}+\mu_{2}^{2})\left[\lambda_{1}\lambda_{2}
K_{21}K_{22}J_{11}I_{12} \right.
\nonumber \\
\left. -\mu_{1}\mu_{2}
K_{11}K_{12}J_{21}I_{22}\right]+(\lambda_{1}^{2}+\mu_{1}^{2})(\lambda_{2}^{2}-\mu_{2}^{2})
\nonumber \\
\times\left[\lambda_{1}\mu_{1}K_{12}K_{21}J_{21}I_{12}-\lambda_{2}\mu_{2}K_{11}K_{22}J_{11}I_{22}\right]
\nonumber \\
+(\lambda_{1}^{2}-\mu_{2}^{2})(\lambda_{2}^{2}+\mu_{1}^{2})\left[\lambda_{1}\mu_{2}K_{12}K_{21}J_{11}I_{22}
\right.
\nonumber \\
\left. -\lambda_{2}\mu_{1}K_{11}K_{22}J_{21}I_{12}\right].
 \label{G}
\end{gather}


\begin{thebibliography}{59}

\bibitem{Jeans1929}
J.~H. Jeans, \emph{Astronomy and
  Cosmogony} (Cambridge University Press, Cambridge, 1929).

\bibitem{Mikhailovskii1977}
  A.~B. Mikhailovskii, V.~I. Petviashvili, and A.~M. Fridman,
  Helical density waves in flat galaxies-moving solitons,
  JETP Lett. \textbf{26}, 121 (1977).

\bibitem{Yueh1981}
  T.~Y. Yueh, Nonlinear waves in a self-gravitating medium,
  Stud. Appl. Math. \textbf{65}, 1 (1981).

\bibitem{Ono1994}
  H.~Ono and I.~Nakata, Soliton formation in a self-gravitating gas,
  Progr. of Theor. Phys. \textbf{92}, 9 (1994).

\bibitem{Zhang1995}
  T.~X. Zhang and X.~Q. Li,  Nonlinear structures of
  self-gravitating systems in stable modes,  Astron. Astrophys. \textbf{294}, 339
  (1995).

\bibitem{Zhang1998}
  H.~Zhang, X.Q. Li, and Y.H. Ma, $N$-soliton pattern in a
  self-gravitating fluid disk,
  Phys. Rev. E \textbf{57}, 1114 (1998).

\bibitem{Gotz1988}
  G.~G\"{o}tz, Solitons in Newtonian gravity,
  Class. Quantum. Grav. \textbf{5}, 743 (1988).

\bibitem{Adams1994}
  F.~C. Adams, M.~Fatuzzo, and R.~Watkins,
  General analytic results for nonlinear waves and solitons in molecular clouds,
  Astrophys. J. \textbf{426}, 629 (1994).

\bibitem{Semelin2001}
  B.~Semelin, N.~S\'{a}nchez, and H.~J. de~Vega,
  Self-gravitating fluid dynamics, instabilities, and solitons,
 Phys. Rev. D \textbf{63}, 084005 (2001).

\bibitem{Verheest1997}
  F.~Verheest and P.~K. Shukla, Nonlinear waves
  in multispecies self-gravitating dusty plasmas,
  Phys. Scr. \textbf{55}, 83 (1997).

\bibitem{Masood2010}
  W.~Masood, H.~A. Shah,  N.~L. Tsintsadze,  and  M.~N.~S. Qureshi,
  Dust Alfv\'{e}n ordinary and  cusp solitons and modulational instability in a self-gravitating
  magneto-radiative plasma,
  Eur. Phys. J. D \textbf{59}, 413 (2010).

\bibitem {Verheest2005}
  T.~Cattaert and F.~Verheest, Solitary waves in self-gravitating molecular clouds,
  Astron. Astrophys. \textbf{438}, 23 (2005).

\bibitem{Fridman1991}
  V.~V. Dolotin and A.~M. Fridman, Generation of
  an observable turbulence spectrum and solitary dipole vortices in rotating
  gravitating systems,
  Sov. Phys. JETP \textbf{72}, 1 (1991).

\bibitem{Jovanovich1990}
  D.~Jovanovi\'{c} and J.~Vranje\v{s}, Vortex solitons in self-gravitating plasma,
  Phys. Scr. \textbf{42},  463 (1990).

\bibitem{Shukla1993}
 P.~K. Shukla, Global vortices in nonuniform gravitating systems,
  Phys. Lett. A \textbf{176}, 54 (1993).

\bibitem{Zinzadze2000}
  N.~L. Tsintsadze, J.~T. Mendonca, P.~K. Shukla, L.~Stenflo,
  and J.~Mahmoodi, Regular structures in self-gravitating dusty plasmas,
  Phys. Scr. \textbf{62}, 70 (2000).

\bibitem{Pokhotelov1998}
  O.~A. Pokhotelov,  V.~V. Khruschev, P.~K. Shukla, L.~Stenflo,
  and J.~F. McKenzie,
  Nonlinearly coupled Rossby-type and inertio-gravity waves in self-gravitating systems,
 Phys. Scr. \textbf{58}, 618 (1998).

\bibitem{Shukla1995}
  P.~K. Shukla and L.~Stenflo, Nonlinear  vortex chains in a nonuniform gravitating fluid,
  Astron. Astrophys. \textbf{300}, 433 (1995).

\bibitem {Abrahamyan2020}
  M.~G. Abrahamyan, Vortices in rotating and gravitating gas disk and in a protoplanetary disk,
  in \emph{Vortex Dynamics Theories and
  Applications}, edited by Z.~Harun (IntechOpen, 2020) pp. 1--21.

\bibitem{Lashkin2023}
V.~M. Lashkin, O.~K. Cheremnykh, Z. Ehsan, and N. Batool,
Three-dimensional vortex dipole solitons in self-gravitating
systems, Phys. Rev. E \textbf{107}, 024201 (2023).

\bibitem{Kuznetsov1986}
E. A. Kuznetsov, A. M. Rubenchik, and V. E. Zakharov, Soliton
stability in plasmas and hydrodynamics, Phys. Rep. \textbf{142},
103-165 (1986).

\bibitem{KuznetsovZakharov2000}
E. A. Kuznetsov and V. E. Zakharov, Nonlinear Coherent Phenomena
in Continuous Media, in \emph{Nonlinear Science at the Dawn of the
21st Century}, Lecture Notes in Physics Vol. 542, edited by P. L.
Christiansen, M. P. S{\o}erensen, and A. C. Scott
(Springer-Verlag, Berlin, 2000), p. 3-46.

\bibitem{Zakharov_UFN2012}
V.~E. Zakharov and E.~A. Kuznetsov, Solitons and collapses: two
evolution scenarios of nonlinear wave systems, Phys.--Usp.
\textbf{55}, 535-556 (2012).

\bibitem{Pedlosky1987}
J. Pedlosky, \emph{Geophysical Fluid Dynamics}  (Springer-Verlag,
New York, 1987).

\bibitem{Charney1948}
J.~G. Charney, On the scale of atmosperic motions, Geophys.
Public. Kosjones Nors. Videnshap.- Akad. Oslo \textbf{17}, 3
(1948).

\bibitem{Hasegawa1978}
A.~Hasegawa and K.~Mima, Pseudo-three dimensional turbulence in
magnetized nonuniform plasma, Phys. Fluids \textbf{21}, 87 (1978).

\bibitem{Larichev1976a}
V.~D. Larichev and G.~M. Reznik, Strongly nonlinear
two-dimensional isolated Rossby waves, Oceanology \textbf{16}, 547
(1976).

\bibitem{Larichev1976b}
V.~D. Larichev and G.~M. Reznik, Two-dimensional Rossby soliton,
an exact solution, Rep. U.S.S.R. Acad. Sci. \textbf{231}, 1077
(1976).

\bibitem{Stern1975}
M. E. Stern, Minimal properties of planetary eddies, J. Mar. Res.
\textbf{33}, 1-13 (1975).

\bibitem{Flierl1987}
G.~R. Flierl, Isolated eddy models in geophysics, Ann. Rev. Fluid
Mech. \textbf{19}, 493 (1987).

\bibitem{Petviashvili_book1992}
O.~A. Pokhotelov and V.~I. Petviashvili, \emph{Solitary Waves in
Plasmas and in the Atmosphere}  (Gordon and Breach, Reading,
1992).

\bibitem{Berestov1979}
A. I. Berestov, Solitary Rossby waves, Izv. Akad. Nauk SSSR Fiz.
Atmos. Okeana \textbf{15}, 648-654 (1979).

\bibitem{Horton1983}
J. D. Meiss and W. Horton, Solitary drift waves in the presence of
magnetic shear, Phys. Fluids \textbf{26}, 990-997 (1983).

\bibitem{Swaters1989}
G. E. Swaters, A perturbation theory for the solitary-drift-vortex
solutions of the Hasegawa-Mima equation, J. Plasma Phys.
\textbf{41}, 523-539 (1989).

\bibitem{Ghil2002}
M. Ghil, Y. Feliks, L. U. Sushama, Baroclinic and barotropic
aspects of the wind-driven ocean circulation, Physica D
\textbf{167}, 1-35 (2002).

\bibitem{Kizner2008}
Z. Kizner, G. Reznik, B. Fridman, R. Khvoles, and J. McWilliams,
Shallow-water modons on the $f$-plane, J. Fluid Mech.
\textbf{603}, 305-329 (2008).

\bibitem{Zeitlin2022}
N. Lahaue and V. Zeitlin, Coherent magnetic modon solutions in
quasi-geostrophic shallow water magnetohydrodynamics, J. Fluid
Mech. \textbf{941}, A15 (2022).

\bibitem{Flierl1980}
G.~R. Flierl, V.~D. Larichev, J.~C. Mc{W}illiams, and G.~M.
Reznik, The dynamics of baroclinic and barotropic solitary eddies,
Dyn. Atmos. Oceans \textbf{5}, 1 (1980).

\bibitem{Makino1981}
M.~Makino, T.~Kamimura, and T.~Taniuti, Dynamics of
two-dimensional   solitary vortices in a low-$\beta$ plasma with
convective motion,  J. Phys. Soc. Jpn. \textbf{50}, 980 (1981).

\bibitem{Williams1982}
J.~C. McWilliams and N.~J. Zabusky, Interactions of isolated
vortices I: Modons colliding with modons, Geophys. Astrophys.
Fluid. Dyn. \textbf{19}, 207 (1982).

\bibitem{Lashkin2017}
V.~M. Lashkin, Stable three-dimensional modon soliton in plasmas,
Phys. Rev. E \textbf{96}, 032211 (2017).

\bibitem{Horton1986}
J. Liu and W. Horton, The intrinsic electromagnetic solitary
vortices in magnetized plasma, J. Plasma Phys. \textbf{36}, 1-24
(1986).

\bibitem{Lakhin1987}
V. P. Lakhin, A. B. Mikha\u{i}lovski\u{i}, and A. I. Smolyakov,
Alfven vortices in a plasma with finite ion temperature, Sov.
Phys. JETP \textbf{65}, 898-903 (1987).

\bibitem{Aburdjania1988PhysScr}
G. D. Aburdzhaniya, Electromagnetic drift vortices in a rotating
plasma cylinder, Phys. Scr. \textbf{38}, 59-63 (1988).

\bibitem{Swaters2004}
G. E. Swaters, Spectral properties in modon stability theory,
Stud. Appl. Math. \textbf{112}, 235-258 (2004).

\bibitem{Nycander1992}
J. Nycander, Refutation of stability proofs for dipole vortices,
Phys. Fluids A \textbf{4}, 467-476 (1992).

\end{thebibliography}
\end{document}